\documentclass[10pt,preprintnumbers,floatfix,letterpaper,aps,prd,nofootinbib,twocolumn,longbibliography,raggedbottom,eqsecnum]{revtex4-1}

\usepackage{amsmath,amssymb,bm}
\usepackage{graphicx}
\usepackage{needspace}
\usepackage{placeins}
\usepackage{xcolor}
\definecolor{revisionred}{RGB}{190,0,0}
\newif\ifshowchanges
\showchangestrue

\usepackage[hypertexnames=false,colorlinks=true,linkcolor=blue,citecolor=blue,urlcolor=blue]{hyperref}
\graphicspath{{figures/}}
\newcommand{\bq}{\begin{equation}}
\newcommand{\eq}{\end{equation}}
\newcommand{\bqn}{\begin{eqnarray}}
\newcommand{\eqn}{\end{eqnarray}}
\newcommand{\nb}{\nonumber}

\begin{document}

\allowdisplaybreaks[1]

\title{First-order corrections to equatorial periodic orbits around rapidly rotating black holes in dynamical Chern--Simons gravity}

\author{Hao-Jie Lin${}^{a,b,c}$}
\email{haojielin@stumail.neu.edu.cn}

\author{Tao Zhu${}^{b,c}$}
\email[Corresponding author: ]{zhut05@zjut.edu.cn}

\author{Jing-Fei Zhang${}^{a}$}
\email{jfzhang@mail.neu.edu.cn}

\author{Xin Zhang${}^{a,d,e}$}
\email[Corresponding author: ]{zhangxin@mail.neu.edu.cn}

\affiliation{
${}^{a}$Key Laboratory of Cosmology and Astrophysics (Liaoning), College of Sciences, Northeastern University, Shenyang 110819, China\\
${}^{b}$ Institute for Theoretical Physics and Cosmology, Zhejiang University of Technology, Hangzhou, 310032, China\\
${}^{c}$ United Center for Gravitational Wave Physics (UCGWP), Zhejiang University of Technology, Hangzhou, 310032, China\\
${}^{d}$ Key Laboratory of Data Analytics and Optimization for Smart Industry (Ministry of Education), Northeastern University, Shenyang 110819, China\\
${}^{e}$ National Frontiers Science Center for Industrial Intelligence and Systems Optimization, Northeastern University, Shenyang 110819, China}

\date{\today}
\begin{abstract}

We calculate first-order corrections to the energy, radial period, and circumferential turning radii of equatorial periodic orbits around rapidly rotating black holes in dynamical Chern--Simons (dCS) gravity. Using an analytic fit to the metric correction, we match each orbit to a Kerr reference at fixed dimensionless spin, dimensionless specific angular momentum, and rational periodicity. The period correction combines a direct contribution at fixed energy with compensation from the energy adjustment that preserves periodicity. Their competition produces sign changes across orbital families and reference paths. In a two-parameter scan, negative period corrections are dominated by energy compensation for sampled prograde orbits and by the direct contribution for sampled retrograde orbits. We locate $C_T=0$ boundaries, where the coefficient of the first-order fractional period correction vanishes. Circumferential shifts can differ in sign from coordinate shifts. Numerical kludge waveforms at leading quadrupole order and their discrete spectra connect these corrections to phase accumulation and harmonic frequency shifts. At fixed dimensionless orbital parameters and physical mass of the small body, this phase contribution can be reproduced to first order by rescaling the Kerr mass. For the $C_T=0$ orbit examined here, this contribution is suppressed at equal black hole masses, while corrections to the waveform and spectral amplitudes remain nonzero.

\end{abstract}

\maketitle

\section{Introduction}
\label{sec:introduction}

Gravitational wave observations provide access to the strong-field dynamics of compact objects~\cite{LIGOScientific:2016aoc,Berti:2015itd,Yunes:2013dva}. LISA will extend these observations to the millihertz band, where extreme-mass-ratio inspirals (EMRIs) are promising sources~\cite{LISA:2024hlh,Babak:2017tow,Gair:2012nm}. In an EMRI, a stellar-mass compact object completes many relativistic orbits around a massive black hole, emitting gravitational waves at harmonics of the orbital frequencies. Small changes in these frequencies can accumulate in the waveform phase over many cycles, allowing the signal to probe the central spacetime geometry and its multipolar structure~\cite{Ryan:1995wh,Barack:2003fp}.

The conservative motion underlying an EMRI can be studied through bound geodesics, which in Kerr spacetime are characterized by fundamental radial, polar, and azimuthal frequencies~\cite{Carter:1968rr,Schmidt:2002qk,Drasco:2003ky,Fujita:2009bp}. Radiation reaction slowly evolves the orbital parameters through much of the inspiral, with self-force methods providing a systematic treatment~\cite{Hughes:1999bq,Poisson:2011nh,Barack:2018yvs}. Near the separatrix, eccentric equatorial trajectories can exhibit zoom--whirl motion, spending several revolutions near periastron before moving outward~\cite{Glampedakis:2002ya,Levin:2008yp,Perez-Giz:2008ajn}. Equatorial periodic orbits organize bound motion through the rational frequency ratio $q=|\Omega_\varphi|/\Omega_r-1=w+v/z$, where the integers $(z,w,v)$ specify the number of leaves, additional whirls, and their visiting order along the direction of motion~\cite{Levin:2008mq,Grossman:2011ps}. The signed azimuthal frequency retains the orbital orientation. A rational label fixes a frequency ratio; the orbital energy, angular momentum, and absolute time scale remain to be specified. Even the full set of fundamental frequencies can be shared by distinct Kerr orbits~\cite{Warburton:2013yj}. Comparisons between different spacetimes therefore require an explicit prescription for selecting and matching the reference orbits.

Periodic orbits have been studied in a variety of black hole backgrounds, including Kehagias--Sfetsos black holes~\cite{Wei:2019zdf}, Einstein--{\AE}ther black holes~\cite{Azreg-Ainou:2020bfl,Lu:2025cxx}, backgrounds with polymer corrections~\cite{Tu:2023xab}, geometries with quantum corrections~\cite{Yang:2024lmj,Zhang:2026ujb}, and black holes in theories with higher curvature interactions~\cite{Lu:2025xlp}. Related studies include magnetically charged regular black holes~\cite{Alloqulov:2025bxh}, black holes in dark matter environments~\cite{Haroon:2025rzx}, and black hole geometries without a Cauchy horizon~\cite{Wang:2025hla}. Dyonic black holes can support distinct periodic branches with the same rational label~\cite{Wang:2025wob}, which can also produce different approximate waveforms~\cite{Wang:2026ncv}. In axisymmetric backgrounds, equatorial periodic motion has been investigated in the static $\gamma$ spacetime~\cite{Zhang:2025wni}, Kerr--Sen black holes~\cite{Liu:2018vea}, rotating black holes with polymer corrections~\cite{Guo:2025scs}, rotating geometries with renormalization group modifications~\cite{Li:2025sfe}, and rotating black holes surrounded by dark matter~\cite{Shahzad:2026hcc}. Further studies address equatorial periodic orbits in rotating Simpson--Visser spacetimes~\cite{Bragado:2025jrg} and rotating black holes inspired by loop quantum gravity~\cite{Yang:2026syx}, as well as resonant bound orbits and their kludge waveforms in rotating Konoplya--Zhidenko spacetimes~\cite{Dai:2026ujj}. Comparisons at fixed angular momentum and rational periodicity, with the energy adjusted accordingly, already have precedents~\cite{Azreg-Ainou:2020bfl,Guo:2025scs,Shahzad:2026hcc,Yang:2026syx}. These comparisons at finite deformation motivate a local perturbative treatment under the same matching conditions. Here we evaluate the constrained coupling derivatives at the Kerr limit to determine the first-order corrections to energy, radial period, and circumferential turning radii.

In dCS gravity, the prescribed scalar of the original Chern--Simons formulation is replaced by a dynamical pseudoscalar coupled to the gravitational Pontryagin density~\cite{Jackiw:2003pm,Alexander:2009tp}. Calculations for compact stars~\cite{Yagi:2013mbt} and observations of neutron stars~\cite{Silva:2020acr} and black holes~\cite{Silva:2022srr} provide complementary information about the coupling under their respective model assumptions. Rotating black hole solutions were first obtained at linear~\cite{Yunes:2009hc} and quadratic~\cite{Yagi:2012ya} orders in spin. Beyond slow rotation, pseudospectral calculations of the scalar field and the trace of the metric perturbation showed that the small-coupling approximation becomes increasingly restrictive at high spin~\cite{Stein:2014xba}. Nonperturbative rotating solutions were also constructed by solving the coupled metric and scalar equations without expanding in either coupling or spin~\cite{Delsate:2018ome}. Spectral methods provide a general framework for constructing stationary, axisymmetric black holes in modified gravity~\cite{Fernandes:2022gde}. Recent spectral constructions yield rapidly rotating dCS metric corrections and analytic fits at leading order in the coupling, without an additional slow-rotation expansion~\cite{Lam:2025elw,Lam:2025fzi}.

Orbital perturbations in deformed Kerr spacetimes have been studied using canonical perturbation theory~\cite{Vigeland:2009pr} and comparisons at fixed turning points~\cite{Gair:2011ym}. In dCS gravity, frequency shifts and waveform dephasing have also been investigated, including waveform comparisons at fixed orbital elements~\cite{Sopuerta:2009iy}. Using the fitted metric, we fix $(a,L,q_*)$, with $L$ the specific angular momentum normalized by the black hole mass, and let the energy and turning points adjust. Differentiating the periodicity constraint determines the energy correction and separates the period correction into direct geometric and energy compensation contributions. We resolve these contributions for the prograde $(3,1,1)$, $(3,2,1)$, and $(3,3,1)$ families. A two-parameter $(3,2,1)$ scan locates the boundaries where the period correction vanishes and compares the mechanisms producing negative period corrections in the sampled prograde and retrograde orbits. The circumferential radius $\mathfrak R=\sqrt{g_{\varphi\varphi}}$ distinguishes geometric changes at the turning points from shifts in their coordinate radii.

We also construct numerical kludge waveforms at leading quadrupole order by mapping the corrected geodesics into an auxiliary flat space~\cite{Babak:2006uv,Sopuerta:2009iy,Canizares:2012is}. Related frameworks include parametrized Kerr metrics~\cite{Vigeland:2011ji,Xin:2018urr} and analytic kludge waveform models~\cite{Gair:2011ym,Chua:2017ujo}. Comparisons at equal normalized closure time and at equal physical time separate periodic waveform corrections from the accumulated phase shift caused by the period change. At fixed dimensionless reference orbit and physical mass $\mu$ of the small body, a Kerr mass rescaling reproduces this common phase contribution to first order. The mass rescaling is related to earlier studies of frequency matching and waveform confusion under parameter adjustments~\cite{Glampedakis:2005cf,Barausse:2006vt}. We also use a representative $C_T=0$ orbit to examine the remaining waveform and spectral corrections at equal mass. Discrete harmonic spectra express the period correction through a common fractional frequency shift, while their amplitude corrections depend on the adopted radiation mapping~\cite{Berry:2012im}.

Section~\ref{section2} introduces the dCS background, equatorial geodesics, and periodicity. Section~\ref{section3} defines the Kerr--dCS matching prescription and develops the first-order derivatives. Section~\ref{section4} specifies the reference orbit paths and presents the orbital corrections, while Sec.~\ref{section5} examines approximate waveforms and spectra. The conclusions and outlook follow in Sec.~\ref{sec:conclusions}.

\section{Rotating black holes and equatorial geodesics in dCS gravity}\label{section2}

The fitted dCS background determines the local geodesic functions for equatorial motion. The radial period and azimuthal advance of bound equatorial orbits
define the rational periodicity used in the Kerr--dCS correspondence below.

\subsection{Background metric and approximation order}\label{sec2:background}

We consider massless, shift-symmetric dCS gravity, in which a pseudoscalar field couples linearly to the gravitational Pontryagin density~\cite{Alexander:2009tp}. In geometrized units, $G=c=1$, the vacuum action is written as~\cite{Lam:2025fzi}
\bqn
S_{\rm dCS}&=&\frac{1}{16\pi}\int d^4x\,\sqrt{-g}\Bigl[\mathcal R-\frac{1}{2}(\nabla\phi)^2 \nb\\
&&\qquad+\alpha\phi\,R_{\mu\nu\rho\sigma}\widetilde R^{\mu\nu\rho\sigma}\Bigr].
\label{dcs_action}
\eqn
Here $\mathcal R$ is the Ricci scalar, and $R_{\mu\nu\rho\sigma}\widetilde R^{\mu\nu\rho\sigma}$ is the Pontryagin density, and $\widetilde R^{\mu\nu\rho\sigma}=\epsilon^{\rho\sigma\alpha\beta}R^{\mu\nu}{}_{\alpha\beta}/2$ is the dual Riemann tensor. The scalar $\phi$ is dimensionless and $\alpha$ has dimensions of length squared. Varying Eq.~\eqref{dcs_action} with respect to the metric and scalar gives the vacuum field equations~\cite{Lam:2025fzi},
\bqn
G_{\mu\nu}+\alpha\mathcal K_{\mu\nu}&=&\frac{1}{2}T^{(\phi)}_{\mu\nu},
\label{sec2:metric_field}\\
\Box\phi&=&-\alpha R_{\mu\nu\rho\sigma}\widetilde R^{\mu\nu\rho\sigma},
\label{sec2:scalar_field}
\eqn
where $G_{\mu\nu}$ is the Einstein tensor, $\Box=\nabla_\mu\nabla^\mu$, and
\bqn
T^{(\phi)}_{\mu\nu}&=&\nabla_\mu\phi\nabla_\nu\phi
-\frac{1}{2}g_{\mu\nu}(\nabla\phi)^2,
\label{sec2:scalar_stress}\\
\mathcal K^{\mu\nu}&=&-4\Bigl[(\nabla_\sigma\phi)
\epsilon^{\sigma\delta\lambda(\mu}\nabla_\lambda R^{\nu)}{}_{\delta}\nb\\
&&\qquad+(\nabla_\sigma\nabla_\delta\phi)
\widetilde R^{\delta(\mu\nu)\sigma}\Bigr].
\label{sec2:cs_tensor}
\eqn
Parentheses denote symmetrization with weight $1/2$. We use signature $(-,+,+,+)$ and $\epsilon^{tr\chi\varphi}=1/\sqrt{-g}$ in the coordinates introduced below. We treat this action as an effective field theory (EFT), retaining the leading metric correction in the small-coupling expansion~\cite{Stein:2014xba}.

For a black hole with ADM mass $M$ and angular momentum $J_{\rm BH}$, we define
\bqn
\zeta&=&\frac{\alpha^2}{M^4}\ll1,\qquad
a=\frac{J_{\rm BH}}{M^2},\qquad |a|<1,
\label{sec2:parameters}
\eqn
where $a$ denotes the dimensionless spin throughout. On the branch continuously connected to Kerr, setting the asymptotic scalar constant to zero gives $\phi=\mathcal O(\alpha/M^2)$. The scalar stress and the Chern--Simons term therefore source the metric at order $\zeta$. At fixed $M$ and $a$, we expand the metric through first order in $\zeta$,
\bqn
g_{\mu\nu}&=&g_{\mu\nu}^{\rm K}+\zeta g_{\mu\nu}^{(1)}+\mathcal O(\zeta^2).
\label{full_metric}
\eqn
At leading order in the coupling, analytic solutions for rotating dCS black holes were obtained at linear~\cite{Yunes:2009hc} and quadratic~\cite{Yagi:2012ya} orders in spin, and subsequently extended to fifth order~\cite{Maselli:2017kic}. Systematic expansions about Kerr allow higher spin orders to be computed~\cite{Cano:2019ore}, with recent constructions obtaining the background metric and scalar field through $\mathcal O(a^{40})$~\cite{Chung:2025gyg}. To study periodic orbits throughout the sampled spin range, we use the analytic metric fits constructed from the spectral solutions of Ref.~\cite{Lam:2025fzi}, without an additional slow-rotation expansion.

For comparison with observational bounds, converting the scalar normalization used in those analyses to that of Eq.~\eqref{dcs_action} multiplies their quoted coupling lengths by $\pi^{1/4}$. The multimessenger neutron star constraint obtained through the I--Love relation therefore corresponds to $\sqrt{|\alpha|}\lesssim\pi^{1/4}(8.5\,\mathrm{km})\simeq11.3\,\mathrm{km}$~\cite{Silva:2020acr}, while the independent black hole ringdown constraint gives $\sqrt{|\alpha|}\lesssim\pi^{1/4}(38.7\,\mathrm{km})\simeq51.5\,\mathrm{km}$~\cite{Silva:2022srr}, both at 90\% credibility. Assuming a universal dCS coupling, these bounds imply $\zeta\lesssim3.45\times10^{-21}$ and $1.48\times10^{-18}$, respectively, for $M=10^6M_\odot$. Their interpretation depends on the adopted stellar or ringdown models and the perturbative approximations used in each analysis. Our primary results are the orbital derivatives evaluated at the Kerr limit; couplings larger than these bounds are used to visualize their first-order dynamical effects.

In Boyer--Lindquist-like coordinates $x^\mu=(t,r,\chi,\varphi)$, with $\chi=\cos\theta$, the metric components through first order in $\zeta$ are
\bqn
g_{tt}&=&-\left(1-\frac{2Mr}{\Sigma}\right)+\zeta H_1, \nb\\
g_{t\varphi}&=&-\frac{2M^2ar}{\Sigma}(1-\chi^2)
\left(1+\zeta H_2\right), \nb\\
g_{rr}&=&\frac{\Sigma}{\Delta}\left(1+\zeta H_3\right), \nb\\
g_{\chi\chi}&=&\frac{\Sigma}{1-\chi^2}
\left(1+\zeta H_3\right), \nb\\
g_{\varphi\varphi}&=&(1-\chi^2)\Bigl[r^2+M^2a^2 \nb\\
&&+\frac{2M^3a^2r}{\Sigma}(1-\chi^2)\Bigr]
\left(1+\zeta H_4\right),
\label{ansatz}
\eqn
where $\Sigma=r^2+M^2a^2\chi^2$, $\Delta=r^2-2Mr+M^2a^2$, and $H_i=H_i(r/M,\chi;a)$ are the fitted correction functions of Ref.~\cite{Lam:2025fzi}, independent of $\zeta$. The mixed term in the line element is $2g_{t\varphi}\,dt\,d\varphi$. At $\zeta=0$, the metric reduces to Kerr, whose horizon radii are $r_\pm=M(1\pm\sqrt{1-a^2})$.

Ref.~\cite{Lam:2025fzi} reports regularized field equation residuals $\lesssim10^{-8}$ for its analytic metric fits at $a\leq0.99$, evaluated by integration over the black hole exterior. The residuals measure how accurately the fitted functions satisfy the field equations, while the validity of the small-coupling expansion at high spin depends on the magnitude of the retained metric deformation~\cite{Stein:2014xba}.

\subsection{Equatorial geodesics and periodicity}\label{sec2:geodesics}

Equatorial reflection symmetry ensures that geodesics initially tangent to $\chi=0$ remain in this plane. All metric components below are evaluated at $\chi=0$. We model the secondary as a structureless, minimally coupled test body, neglecting its spin and scalar multipole interactions~\cite{Yagi:2013mbt}, as well as self-force and radiation reaction~\cite{Sopuerta:2009iy}.

With $u^\mu=dx^\mu/d\tau$ the four-velocity, stationarity and axisymmetry provide the conserved specific energy $E=-u_t$ and dimensionless specific azimuthal angular momentum $L=u_\varphi/M$. An overdot denotes differentiation with respect to the proper time $\tau$. Solving for $\dot t$ and $\dot\varphi$ gives
\bqn
U\equiv\dot t&=&\frac{g_{\varphi\varphi}E+M g_{t\varphi}L}
{g_{t\varphi}^{2}-g_{tt}g_{\varphi\varphi}},\label{sec2:U}\\
\Phi\equiv\dot\varphi&=&-\frac{g_{t\varphi}E+M g_{tt}L}
{g_{t\varphi}^{2}-g_{tt}g_{\varphi\varphi}},\label{sec2:Phi}
\eqn
and the normalization $u^\mu u_\mu=-1$ yields
\bqn
R\equiv\dot r^2&=&-\frac{1}{g_{rr}} \nb\\
&&+\frac{g_{\varphi\varphi}E^2+2M g_{t\varphi}EL+M^2g_{tt}L^2}
{g_{rr}(g_{t\varphi}^{2}-g_{tt}g_{\varphi\varphi})}.
\label{sec2:R}
\eqn
The functions $U$, $\Phi$, and $R$ provide the local inputs from the metric needed for the equatorial motion below. We set $M=1$ in the numerical geodesic calculations.

For a bound exterior orbit, the periastron $r_p$ and apastron $r_a$ are distinct simple roots of $R$, with $R>0$ between them. The coordinate time elapsed and azimuthal angle accumulated over one complete radial oscillation are
\bqn
T_r&=&2\int_{r_p}^{r_a}\frac{U}{\sqrt R}\,dr,\label{sec2:Tr}\\
\Delta\varphi&=&2\int_{r_p}^{r_a}\frac{\Phi}{\sqrt R}\,dr,\label{sec2:delta_phi}
\eqn
which define the fundamental frequencies with respect to the coordinate time,
\bqn
\Omega_r&=&\frac{2\pi}{T_r},\qquad
\Omega_\varphi=\frac{\Delta\varphi}{T_r}.
\label{sec2:frequencies}
\eqn
In the Kerr limit, these quantities coincide with the standard radial and azimuthal frequencies of bound equatorial geodesics~\cite{Schmidt:2002qk,Drasco:2003ky,Fujita:2009bp}.

For periodic motion in either orientation, we use the rational classification of Ref.~\cite{Levin:2008mq},
\bqn
q&\equiv&\frac{|\Omega_\varphi|}{\Omega_r}-1
=\frac{|\Delta\varphi|}{2\pi}-1=w+\frac vz.
\label{sec2:q}
\eqn
Here $z$ is the number of leaves, equivalently the number of radial oscillations required for closure, $w$ counts the additional complete whirls per radial oscillation, and $v$ specifies their visiting order along the direction of motion. The fraction $v/z$ is written in lowest terms, and the closure time is $T_{\rm cl}=zT_r$. Fixing $q$ specifies the frequency ratio, while the energy, turning radii, and radial period depend on the background geometry and the chosen orbit correspondence. We restrict the differentiation to nondegenerate bound orbits. In the circular limit, the two turning points merge; at the eccentric Kerr separatrix, the periastron becomes a double radial root associated with an unstable circular orbit~\cite{Levin:2008yp}.

We take $a>0$, with $L>0$ denoting prograde motion and $L<0$ retrograde motion. The signs of $\Phi$, $\Delta\varphi$, and $\Omega_\varphi$ retain the orientation, while Eq.~\eqref{sec2:q} uses the magnitudes of the frequency and the azimuthal advance.

\section{First-order construction at fixed periodicity}\label{section3}

Our Kerr--dCS correspondence holds the dimensionless spin $a$, specific angular momentum $L$, and rational periodicity $q_*$ fixed as the coupling varies. The energy adjusts to preserve $q_*$, and the turning points and radial period change accordingly. This prescription preserves the periodic label; alternative matching conditions select different orbital families. Related studies treat perturbative modifications of strong-field geodesics in dCS and parametrically deformed Kerr spacetimes~\cite{Sopuerta:2009iy,Vigeland:2011ji,Gair:2011ym}. We evaluate the derivatives of this family with respect to $\zeta$ at the Kerr limit. For any orbital quantity $X$ we write
\bqn
X(\zeta)&=&X^{\rm K}+\zeta\,\delta X+\mathcal O(\zeta^2),
\label{sec3w:expansion}
\eqn
Here $\delta X$ denotes the coupling derivative along the family with $(a,L,q_*)$ fixed; the displayed first-order shift is $\zeta\delta X$. The partial derivatives used below hold $(a,L)$ fixed: $\partial/\partial\zeta$ also fixes $E$, whereas $\partial/\partial E$ fixes $\zeta$. They are evaluated at the Kerr orbit after differentiation. The construction applies to smooth branches of exterior bound orbits with distinct turning points that are simple roots of $R$. We also require $\partial q/\partial E\neq0$.

We first identify the Kerr orbit with the prescribed $(a,L,q_*)$ by solving $q(E^{\rm K},0)=q_*$. Its periastron and apastron set the limits of the radial integrals. The regularized integrals below give the azimuthal advance, the radial period, and the changes caused by varying the energy or coupling. We evaluate the local derivatives analytically and the integrals by Gauss--Legendre quadrature. These results determine the energy adjustment $\delta E$ needed to preserve $q_*$. We then include this adjustment in the period and turning point shifts and reconstruct the motion along the orbit.

\subsection{Regularized orbital derivatives}\label{sec3w:integrals}

For $F=U,\Phi$, define the integral over one radial oscillation
\bqn
I_F&=&2\int_{r_p}^{r_a}\frac{F}{\sqrt R}\,dr,\nb\\
T_r&=&I_U,
\qquad q=\frac{|I_\Phi|}{2\pi}-1.
\label{sec3w:IF}
\eqn
Varying $E$ or $\zeta$ shifts the turning points. Differentiating the root condition gives
\bqn
\frac{\partial r_i}{\partial s}
=-\left.\frac{\partial R/\partial s}{\partial R/\partial r}\right|_{r=r_i}, \quad i=p,a,\quad s=E,\zeta.
\label{sec3w:root}
\eqn
In Eq.~\eqref{sec3w:IF}, the integration limits depend on the parameters and the integrand has integrable singularities at the turning points. We map the interval to $0\le\eta\le\pi$,
\bqn
r(\eta)&=&\frac{r_a+r_p}{2}-\frac{r_a-r_p}{2}\cos\eta, \nb\\*
W(\eta)&\equiv&\frac{d\tau}{d\eta}
=\frac{(r_a-r_p)\sin\eta}{2\sqrt{R}}, \nb\\*
I_F&=&2\int_0^\pi WF\,d\eta.
\label{sec3w:map}
\eqn
This type of radial reparameterization is standard in Kerr orbital integrals~\cite{Drasco:2003ky,Fujita:2009bp}. On the fixed interval, the chain rule gives
\bqn
\frac{\partial I_F}{\partial s}
&=&2\int_0^\pi\!\Biggl[
W\biggl(\frac{\partial F}{\partial s}
+\frac{\partial F}{\partial r}\left.\frac{\partial r}{\partial s}\right|_\eta\biggr)\nb\\*
&&\qquad{}+F\left.\frac{\partial W}{\partial s}\right|_\eta
\Biggr]d\eta.
\label{sec3w:derivative}
\eqn
The three terms describe the local change of $F$, the displacement of the sampled radius, and the change of the weight $W$.

At a simple turning point, the particle reverses its radial motion in finite proper time. Writing $R=(r-r_p)(r_a-r)\mathcal S$, Eq.~\eqref{sec3w:map} gives
\bqn
\frac{dr}{d\eta}&=&\frac{r_a-r_p}{2}\sin\eta, \nb\\
(r-r_p)(r_a-r)&=&\frac{(r_a-r_p)^2}{4}\sin^2\eta. 
\eqn
Since $\sin\eta>0$ on $0<\eta<\pi$, the endpoint factors cancel:
\bqn
W=\frac{d\tau}{d\eta}&=&\frac{dr/d\eta}{\sqrt R}
=[\mathcal S(r(\eta);E,\zeta)]^{-1/2}. 
\eqn
For a smooth family with distinct simple roots and $R>0$ between them, $\mathcal S$ remains positive and bounded away from zero near the reference orbit. Thus $WU$, $W\Phi$, and their first parameter derivatives at fixed $\eta$ have finite endpoint limits. The parameter dependence of $r(\eta)$ and $W$ already includes the endpoint motion, so no separate boundary term is needed. The local derivatives follow from Eqs.~\eqref{sec2:U}--\eqref{sec2:R}.

\subsection{Corrections at fixed periodicity}\label{sec3w:corrections}

The derivatives along the matched family are obtained by imposing $q=q_*$. The reference Kerr energy satisfies $q(E^{\rm K},0)=q_*$. Expanding this constraint,
\bqn
q(E^{\rm K}+\zeta\delta E,\zeta)
=q_*+\zeta\Bigl(\frac{\partial q}{\partial\zeta} +\delta E\frac{\partial q}{\partial E}\Bigr)
+\mathcal O(\zeta^2),\nb\\
\label{sec3w:qexpand}
\eqn
therefore gives
\bqn
\delta E&=&-\frac{\partial q/\partial\zeta}{\partial q/\partial E}, \nb\\
\frac{\partial q}{\partial s}&=&\frac{1}{2\pi}\frac{\partial|I_\Phi|}{\partial s}.
\label{sec3w:energy}
\eqn
The numerator measures the direct coupling correction at fixed energy, and the denominator is the Kerr energy sensitivity. The derivative of $|I_\Phi|$ is evaluated on a fixed-orientation branch with $I_\Phi\ne0$, so $\partial_s|I_\Phi|=\operatorname{sgn}(I_\Phi)\,\partial_sI_\Phi$.

Once $\delta E$ is fixed, any smooth orbital scalar $X(E,\zeta)$ acquires the constrained coefficient
\bqn
\delta X&=&\left.\frac{\partial X}{\partial\zeta}\right|_{E}
+\delta E\left.\frac{\partial X}{\partial E}\right|_{\zeta=0}.
\label{sec3w:total_correction}
\eqn
In particular,
\bqn
\delta r_i&=&-\left.
\frac{\partial R/\partial\zeta+\delta E\,\partial R/\partial E}
{\partial R/\partial r}\right|_{r=r_i^{\rm K}},\quad i=p,a, \nb\\*
\delta T_r&=&\frac{\partial I_U}{\partial\zeta}
+\delta E\frac{\partial I_U}{\partial E}.
\label{sec3w:responses_eq}
\eqn
On a fixed-orientation branch, preserving $q$ gives $\delta(\Delta\varphi)=0$. Hence
\bqn
\frac{\delta\Omega_r}{\Omega_r^{\rm K}}
&=&\frac{\delta\Omega_\varphi}{\Omega_\varphi^{\rm K}}
=-\frac{\delta T_r}{T_r^{\rm K}}.
\label{sec3w:frequency_relation}
\eqn
The common fractional frequency shift is a kinematic consequence of fixing $q_*$. The matching preserves the azimuthal advance per radial cycle while allowing its absolute time scale to change. This difference underlies the waveform analysis in Sec.~\ref{section5}.

\subsection{Trajectory reconstruction and time normalization}\label{sec3w:trajectory}

To compare the closed trajectories, we start both orbits at periastron with $t(0)=\varphi(0)=0$. As $\eta$ increases from $0$ to $2\pi z$, the elapsed coordinate time and accumulated azimuthal angle are
\bqn
t(\eta)&=&\int_0^\eta W(\eta')U(\eta')\,d\eta', \nb\\
\varphi(\eta)&=&\int_0^\eta W(\eta')\Phi(\eta')\,d\eta'. 
\eqn
The positive weight $W=|dr/d\eta|/\sqrt R$ keeps proper time increasing on both the outward and inward legs and remains finite at the turning points. At fixed $\eta$, the first-order changes in these integrals include the metric deformation, the energy adjustment needed to preserve $q_*$, and the shifts of the turning points [Eq.~\eqref{sec3w:derivative}]. We then use $t(\eta)$ to compare positions at the same fraction $u$ of each orbit's closure period.

For the orbit figures we use normalized closure time,
\bqn
u&\equiv&\frac{t}{T_{\rm cl}(\zeta)},\qquad T_{\rm cl}=zT_r.
\eqn
Expanding $t(\eta;\zeta)=uT_{\rm cl}(\zeta)$ gives the shift in $\eta$ required for comparison at fixed $u$,
\bqn
\left.\delta\eta\right|_u
&=&\frac{uz\,\delta T_r-\left.\delta t\right|_\eta}
{W^{\rm K}U^{\rm K}}, \nb\\
\left.\delta X\right|_u
&=&\left.\delta X\right|_\eta+\frac{dX^{\rm K}}{d\eta}\left.\delta\eta\right|_u,
\quad X=r,\varphi.
\label{sec3w:clock}
\eqn
Equal $u$ pairs corresponding fractions of each model's closure time; their physical times generally differ. At common $u$, the auxiliary Cartesian embedding $x=r\cos\varphi$, $y=r\sin\varphi$ has
\bqn
\delta x&=&\delta r\cos\varphi^{\rm K}
-r^{\rm K}\delta\varphi\sin\varphi^{\rm K}, \nb\\
\delta y&=&\delta r\sin\varphi^{\rm K}
+r^{\rm K}\delta\varphi\cos\varphi^{\rm K}.
\label{sec3w:cartesian}
\eqn
The aligned $r(u)$ and $\varphi(u)$ determine both the representation using the circumferential radius in Sec.~\ref{section4} and the coordinate embedding used for the waveforms in Sec.~\ref{section5}. The former includes the local metric correction when converting $r$ to $\mathfrak{R}$; the latter uses $x=r\cos\varphi$ and $y=r\sin\varphi$ as specified above. A comparison at the same physical coordinate time is different because the correction to $T_{\rm cl}$ must then remain explicit; Sec.~\ref{section5} carries out that conversion at waveform level.

\section{First-order orbital dynamics}\label{section4}

We specify the Kerr path in angular momentum for either orientation by a single dimensionless fraction,
\bqn
f_L&\equiv&\frac{|L|-|L_{\rm ISCO}^{\rm K}|}
{|L_{\rm MBO}^{\rm K}|-|L_{\rm ISCO}^{\rm K}|}.
\label{sec4:Lpath}
\eqn
Here $L_{\rm ISCO}^{\rm K}$ and $L_{\rm MBO}^{\rm K}$ are the dimensionless specific angular momenta of the Kerr ISCO and MBO for each orientation~\cite{Bardeen:1972fi}. Fixing $f_L$ determines $|L(a)|$ along a spin sequence; the orientation fixes its sign. The prograde baseline is anchored at $a_0=0.5$ by $L(a_0)=3.3$, giving $f_L\simeq0.776642$. Along this path, we study $0.5\leq a\leq0.99$ for the $(3,1,1)$, $(3,2,1)$, and $(3,3,1)$ families, corresponding to $q_*=4/3$, $7/3$, and $10/3$, respectively. The coupling derivatives are evaluated at $\zeta=0$, and finite first-order shifts are displayed at $\zeta=10^{-5}$ unless stated otherwise.

Changing $a$ along a spin sequence changes $L(a)$, whereas each coupling derivative keeps the corresponding reference values of $(a,L,q_*)$ fixed. Unless stated otherwise, the prograde spin sequence uses the baseline value of $f_L$.

\subsection{Direct geometric corrections and energy compensation}\label{sec4:precession}

\begin{figure*}[!t]
\centering
\includegraphics[width=0.96\textwidth]{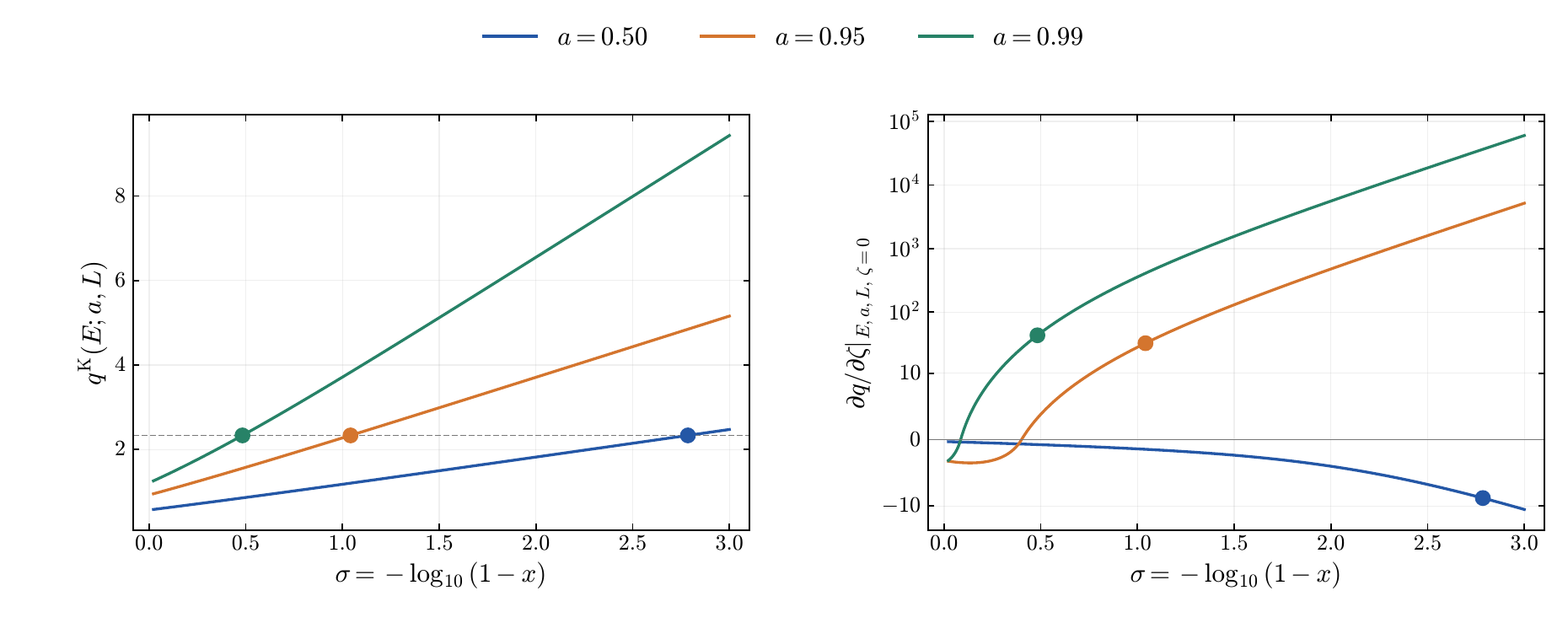}
\caption{Matching of prograde periodic orbits for $a=0.50,0.95,0.99$ on the baseline path in angular momentum. The left plot selects $q_*=7/3$ on the Kerr background. The right plot shows the direct derivative $\partial q/\partial\zeta$ at fixed $E,a,L$, before the compensating energy correction is included. The horizontal coordinate is the transformed energy fraction defined in Sec.~\ref{sec4:precession}; the logarithmic display in the right plot preserves the sign, and its ticks give the untransformed derivative.}
\label{fig1_periodicity}
\end{figure*}

The left plot of Fig.~\ref{fig1_periodicity} illustrates the selection of a prograde Kerr reference by $q^{\rm K}(E)=q_*=7/3$, corresponding to $(z,w,v)=(3,2,1)$. At fixed $(a,L)$, the bound orbit interval is delimited by the energies $E_{\min}^{\rm K}$ and $E_{\max}^{\rm K}$ of the stable and unstable circular orbits. Both satisfy $R^{\rm K}=\partial_rR^{\rm K}=0$, with stability distinguished by $\partial_r^2R^{\rm K}<0$ and $>0$, respectively. For the adopted prograde range $L_{\rm ISCO}^{\rm K}<L<L_{\rm MBO}^{\rm K}$, these endpoints obey $0<E_{\min}^{\rm K}<E_{\max}^{\rm K}<1$ and are evaluated at the selected angular momentum. The plotting coordinate $\sigma=-\log_{10}(1-x)$, with $x=(E-E_{\min}^{\rm K})/(E_{\max}^{\rm K}-E_{\min}^{\rm K})$, magnifies the upper part of this interval.

The direct periodicity derivative $\partial q/\partial\zeta$ is evaluated at fixed $E,a,L$, before imposing the matching constraint [the right plot of Fig.~\ref{fig1_periodicity}]. At the selected orbits it is approximately $-7.26994$, $31.7068$, and $42.5627$ for $a=0.50,0.95,0.99$, respectively. The Kerr slope $q_{,E}$ is positive at these references, so the energy correction has the opposite sign to the direct periodicity response. Equation~\eqref{sec3w:energy} gives
\bqn
\delta E&\simeq&0.00153225,\;-0.336723,\;-1.12027.
\label{sec4:eq_energy_values}
\eqn
These energy adjustments preserve $q_*$ to first order.

Along the family at fixed periodicity, $T_r=T_r(E(\zeta),\zeta)$. Applying Eq.~\eqref{sec3w:total_correction} and normalizing by the Kerr period gives
\bqn
C_T&\equiv&\frac{\delta T_r}{T_r^{\rm K}}
=C_T^{\mathrm{dir}}+C_T^{\mathrm{comp}},\nb\\
C_T^{\mathrm{dir}}&\equiv&
\frac{1}{T_r^{\rm K}}\frac{\partial T_r}{\partial\zeta},\nb\\
C_T^{\mathrm{comp}}&\equiv&
\frac{\delta E}{T_r^{\rm K}}\frac{\partial T_r}{\partial E}.
\label{sec4:period_split}
\eqn
Here \mbox{$C_T^{\mathrm{dir}}$} is the direct contribution at fixed $E,a,L$, and \mbox{$C_T^{\mathrm{comp}}$} is the energy compensation required to preserve $q_*$. The energy derivative holds $(a,L,\zeta)$ fixed. Both derivatives are evaluated at the Kerr reference and include the induced endpoint motion. Their competition determines the finite fractional change, $[T_r(\zeta)-T_r^{\rm K}]/T_r^{\rm K}=\zeta C_T+\mathcal O(\zeta^2)$.

\subsection{Spin dependence and circumferential radii}\label{sec4:spin_labels}

The geometric scale of each radial endpoint is characterized by its circumferential radius. With the azimuthal angle normalized to period $2\pi$, an equatorial circle at fixed $t$ and $r$ has proper circumference $C=2\pi\sqrt{g_{\varphi\varphi}}$. We therefore define
\bqn
\mathfrak R(r;\zeta)&\equiv&\frac{C}{2\pi}
=\sqrt{g_{\varphi\varphi}(r;\zeta)}.
\label{sec4:eq_circumference}
\eqn
This perimetral radius~\cite{Fernandes:2022gde} measures the size of an axial circle and is unchanged by a smooth relabeling of the radial coordinate. At a turning point $i=p,a$, differentiating $\mathfrak R_i(\zeta)=\sqrt{g_{\varphi\varphi}(r_i(\zeta);\zeta)}$ at $\zeta=0$ and dividing by $\mathfrak R_i^{\rm K}$ gives
\bqn
\frac{\delta\mathfrak R_i}{\mathfrak R_i^{\rm K}}
&=&\left.\frac{g_{\varphi\varphi}^{(1)}
+\partial_rg_{\varphi\varphi}^{\rm K}\,\delta r_i}
{2g_{\varphi\varphi}^{\rm K}}\right|_{r=r_i^{\rm K}}.
\label{sec4:eq_circumference_response}
\eqn
Here $\delta r_i$ is the correction at fixed $(a,L,q_*)$ obtained in Sec.~\ref{section3}. The two terms transform together under a radial coordinate change; their sum gives the change of circumference.

\begin{figure*}[!t]
\centering
\includegraphics[width=\textwidth]{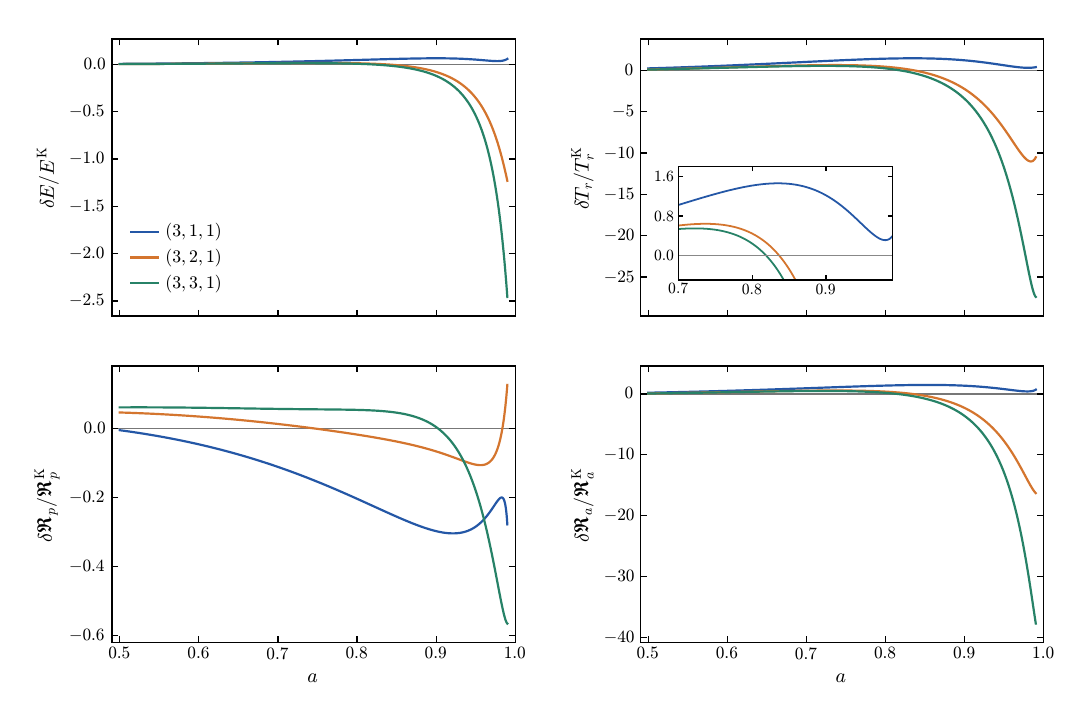}
\caption{First-order correction coefficients for three prograde periodic families along the baseline $f_L=0.776642$ spin sequence. The upper row shows energy (left) and radial period (right); the lower row shows the circumferential periastron (left) and apastron (right), including metric and turning point contributions [Eq.~\eqref{sec4:eq_circumference_response}]. The shared legend in the upper-left plot identifies $(z,w,v)$, with $q_*=4/3,7/3,10/3$. The inset in the upper-right plot resolves the period response near zero. All vertical axes show coefficients per unit $\zeta$, evaluated at $\zeta=0$.}
\label{fig2_spin_corrections}
\end{figure*}

Figure~\ref{fig2_spin_corrections} compares the coefficients for energy, radial period, and circumferential turning radii for the matched families $(3,1,1)$, $(3,2,1)$, and $(3,3,1)$ over $0.50\leq a\leq0.99$. The plotted ratios $\delta X/X^{\rm K}$ are coefficients per unit coupling; the actual first-order fractional changes are $\zeta\delta X/X^{\rm K}$. Both the magnitude and sign of the energy and period coefficients depend strongly on the periodic label. At $a=0.99$, in the order $(3,1,1)$, $(3,2,1)$, $(3,3,1)$,
\bqn
\frac{\delta E}{E^{\rm K}}&\simeq&(0.056576,-1.230516,-2.457174),
\label{sec4:eq_highspin_energy}\\
\frac{\delta T_r}{T_r^{\rm K}}&\simeq&(0.390218,-10.518613,-27.403328).
\label{sec4:eq_highspin_values}
\eqn
For $(3,1,1)$ at $a=0.99$, the direct contribution is $0.004380$, while energy compensation contributes $0.385838$ and supplies most of $C_T$. For $(3,2,1)$ and $(3,3,1)$, negative compensation exceeds the positive direct term: $4.715079-15.233692=-10.518613$ and $17.121740-44.525068=-27.403328$. At these reference points $T_{r,E}>0$, so lowering the energy shortens the period. Equation~\eqref{sec3w:frequency_relation} relates this change to the common fractional frequency correction and the waveform phase accumulation in Sec.~\ref{section5}.

The circumferential turning radii provide a complementary geometric comparison. For $(3,2,1)$ at $a=0.99$,
\bqn
\frac{\delta\mathfrak R_p}{\mathfrak R_p^{\rm K}}&\simeq&+0.126418,
\label{sec4:eq_circumference_periastron}\\
\frac{\delta\mathfrak R_a}{\mathfrak R_a^{\rm K}}&\simeq&-16.2952.
\label{sec4:eq_circumference_values}
\eqn
At $\zeta=10^{-5}$, the periastron circumference increases by about $0.0001264\%$, whereas the apastron circumference decreases by $0.01630\%$. The two negative coordinate shifts $\delta r_p$ and $\delta r_a$ in the fitted radial gauge therefore do not imply that both circumferences decrease. Opposite signs of coordinate and circumferential shifts also occur for circular orbits in the example from modified gravity in Ref.~\cite{Fernandes:2022gde}.

At $\zeta=10^{-5}$, the largest absolute fractional correction among $E$, $T_r$, $\mathfrak R_p$, and $\mathfrak R_a$ in the sampled baseline orbits is $3.77\times10^{-4}$, in the $(3,3,1)$ apastron circumferential radius at $a=0.99$. These changes remain locally small, with pronounced dependence on spin and periodic family at high spin.

\subsection{Prograde and retrograde $C_T=0$ boundaries}\label{sec4:cancellation}

The primary grid comprises 5360 Kerr reference orbits whose radial turning points are simple roots. They span 82 common spins, with $f_{\min}(a)<f_L<1$ for each orientation. Here $f_{\min}(a)$ is the circular limit of the $(3,2,1)$ family; samples stay a fraction $10^{-8}$ of this band width from each endpoint. Prograde and retrograde motion use their respective Kerr ISCO--MBO bands, so equal $f_L$ generally selects different reference geometries.

\begin{figure*}[!t]
\centering
\includegraphics[width=\textwidth]{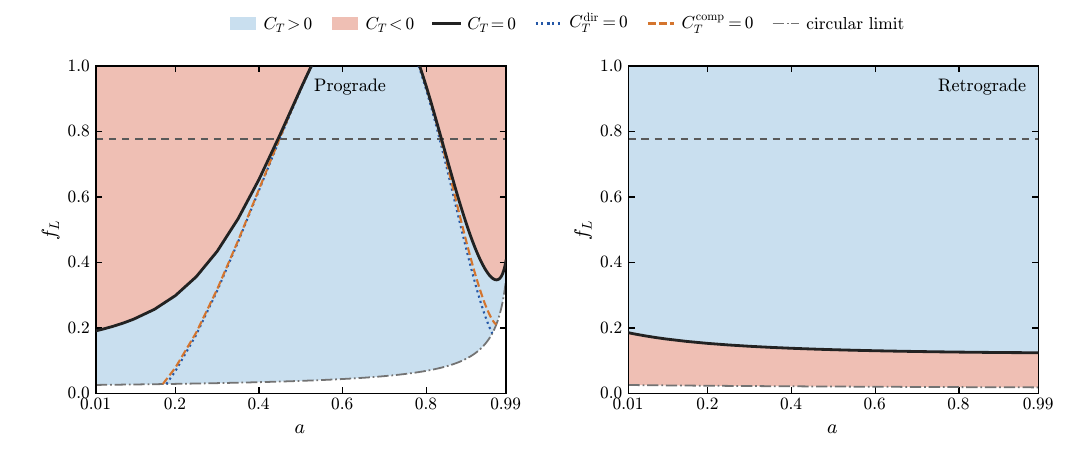}
\caption{Period response of the prograde and retrograde $(3,2,1)$ families at fixed $(a,L,q_*)$, with $q_*=7/3$. The two plots show the sign regions in $(a,f_L)$ over $0.01\leq a\leq0.99$, for prograde (left) and retrograde (right) motion, using each orientation's Kerr ISCO--MBO band. The coefficient is $C_T=C_T^{\mathrm{dir}}+C_T^{\mathrm{comp}}$, where $C_T^{\mathrm{dir}}$ is the direct contribution and $C_T^{\mathrm{comp}}$ is the contribution from energy compensation [Eq.~\eqref{sec4:period_split}]. Black solid, blue dotted, and orange dashed curves join zeros of $C_T$, $C_T^{\mathrm{dir}}$, and $C_T^{\mathrm{comp}}$, respectively. The lower boundary is the circular limit; gray dashed lines mark $f_L=0.776642$. All coefficients are evaluated at $\zeta=0$.}
\label{fig3_cancellation_map}
\end{figure*}

Figure~\ref{fig3_cancellation_map} maps the $(3,2,1)$ period response over $0.01\leq a\leq0.99$. The black $C_T=0$ curves bound negative prograde responses on the side with larger $f_L$ and negative retrograde responses on the side with smaller $f_L$. The retrograde $C_T=0$ boundary extends across the sampled spin range. The prograde $C_T=0$ boundary has two segments meeting the upper edge of the selected band, separated by an interval in which the sampled family has $C_T>0$ throughout. The sign along a spin sequence therefore depends on the chosen fixed $f_L$.

\begin{figure}[!tb]
\centering
\includegraphics[width=\columnwidth]{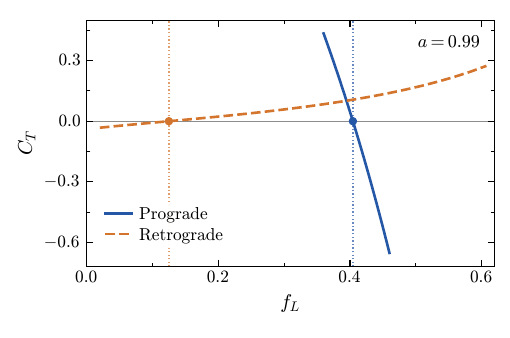}
\caption{Period response coefficient $C_T$ for the prograde and retrograde $(3,2,1)$ families at $a=0.99$, evaluated at fixed $(a,L,q_*)$, with $q_*=7/3$. Blue solid and orange dashed curves compare prograde and retrograde motion on common axes. The displayed intervals are $0.36\leq f_L\leq0.46$ for prograde motion and approximately $0.02\lesssim f_L\lesssim0.61$ for retrograde motion. Dots and vertical dotted lines mark $C_T=0$, crossed with opposite slopes in the two orientations. All coefficients are evaluated at $\zeta=0$.}
\label{fig4_period_response}
\end{figure}

At $a=0.99$, Fig.~\ref{fig4_period_response} resolves the opposite crossings. The dots mark the numerically located zeros $f_L^*\simeq0.404534$ and $0.124883$ for prograde and retrograde motion, respectively. Increasing $f_L$ crosses from positive to negative $C_T$ for prograde motion and from negative to positive for retrograde motion.

For sampled $C_T<0$ references, the evaluated period decomposition [Eq.~\eqref{sec4:period_split}] gives \mbox{$C_T^{\mathrm{dir}}>0$} and \mbox{$C_T^{\mathrm{comp}}<-C_T^{\mathrm{dir}}$} for prograde motion, and \mbox{$C_T^{\mathrm{dir}}<0$} and \mbox{$0<C_T^{\mathrm{comp}}<|C_T^{\mathrm{dir}}|$} for retrograde motion. Negative energy compensation therefore dominates prograde shortening, whereas the negative direct term dominates retrograde shortening. At the retrograde $C_T=0$ root for $a=0.99$, \mbox{$C_T^{\mathrm{dir}}\simeq-5.136199$} and \mbox{$C_T^{\mathrm{comp}}\simeq5.136199$}: the nonzero contributions cancel. The prograde $C_T=0$ orbit will provide the corresponding waveform comparison in Sec.~\ref{sec5:clocks}.

\subsection{Representative orbit diagrams}\label{sec4:deformation}

\begin{figure*}[!t]
\centering
\includegraphics[width=\textwidth]{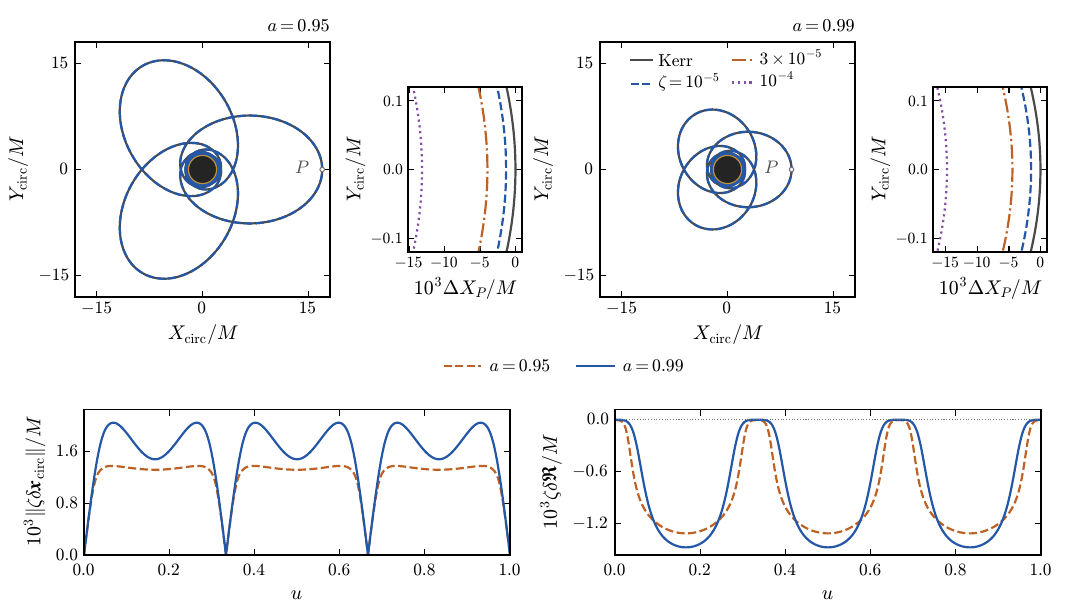}
\caption{Baseline prograde $(3,2,1)$ orbits in circumferential radius at equal normalized closure time. Top: $a=0.95$ (left) and $0.99$ (right), with adjacent apastron enlargements. Full orbits compare Kerr and first-order dCS at $\zeta=10^{-5}$ on equal axis scales. Enlargements rescale the same reconstruction to $\zeta=10^{-5},3\times10^{-5},10^{-4}$, with $\Delta X_P=X_{\rm circ}-\mathfrak R_a^{\rm K}$. Disks mark the Kerr horizon reference $\mathfrak R_H^{\rm K}=2M$. Bottom: planar displacement magnitude (left) and circumferential radius change $\zeta\delta\mathfrak R(u)$ (right), both at $\zeta=10^{-5}$.}
\label{fig5_periodic_orbits}
\end{figure*}

The $(3,2,1)$ orbits at $a=0.95$ and $0.99$ are compared in Fig.~\ref{fig5_periodic_orbits} at the same normalized closure time $u$ defined in Sec.~\ref{section3}. Each closure retains three radial oscillations and ten azimuthal revolutions. The radial coordinate of the diagram is $\mathfrak R$, so that each plotted axial circle has the corresponding geometric circumference.

Along the Kerr trajectory, $\delta\mathfrak R|_u$ is obtained from Eq.~\eqref{sec4:eq_circumference_response} by replacing $r_i^{\rm K}$ and $\delta r_i$ with $r^{\rm K}(u)$ and $\delta r|_u$. The pointwise planar correction is
\bqn
&&\boldsymbol{x}_{\rm circ}
=\mathfrak R(\cos\varphi,\sin\varphi),\\
&&\left.\delta\boldsymbol{x}_{\rm circ}\right|_u
=\left.\delta\mathfrak R\right|_u\,\boldsymbol e_r
+\mathfrak R^{\rm K}\left.\delta\varphi\right|_u\,\boldsymbol e_\varphi,
\label{sec4:eq_circumference_orbit}\\
&&\frac{\boldsymbol{x}_{\rm circ}^{\rm K}\cdot
\zeta\delta\boldsymbol{x}_{\rm circ}}{\mathfrak R^{\rm K}}
=\zeta\left.\delta\mathfrak R\right|_u.
\label{sec4:eq_radial_projection}
\eqn
The unit vectors $\boldsymbol e_r=(\cos\varphi^{\rm K},\sin\varphi^{\rm K})$ and $\boldsymbol e_\varphi=(-\sin\varphi^{\rm K},\cos\varphi^{\rm K})$ are evaluated at the Kerr azimuth $\varphi^{\rm K}(u)$. Equations~\eqref{sec4:eq_circumference_orbit} and~\eqref{sec4:eq_radial_projection} follow by differentiation at fixed $u$ and projection onto the Kerr radial direction, respectively. Both the radial and azimuthal corrections enter the displacement magnitude in the lower-left plot; the lower-right plot isolates the change of circumferential radius.

For $a=0.95$, the correction to the circumferential radius is negative throughout the sampled closure. At $a=0.99$, it is positive near periastron and negative at apastron, consistent with Eqs.~\eqref{sec4:eq_circumference_periastron} and~\eqref{sec4:eq_circumference_values}. At $\zeta=10^{-5}$, the maximum displacements in the sampled orbit diagrams are approximately $1.378\times10^{-3}M$ and $2.046\times10^{-3}M$ for the two spins, respectively, and occur away from the turning points.

\section{Approximate gravitational waveforms and spectra}\label{section5}

All waveforms and spectra in this section use the prograde branch. The corrected conservative motion produces characteristic zoom--whirl waveform structure~\cite{Glampedakis:2002ya}. Comparisons at equal normalized closure time and at equal physical time separate the periodic waveform correction from the phase drift accumulated through the period shift.

\subsection{Quadrupole waveforms and time conventions}\label{sec5:model}

The equatorial trajectory is mapped to auxiliary Cartesian coordinates,
\bqn
\boldsymbol X_p(t)&=&(r\cos\varphi,r\sin\varphi,0), \nb\\
V_i&=&\frac{dX_i}{dt},\qquad A_i=\frac{d^2X_i}{dt^2}. \label{sec5:embedding}
\eqn
The orbit diagrams in Sec.~\ref{section4} use the circumferential radius $\mathfrak{R}$ to describe the geometry of axial circles. For waveform construction, the same geodesics are mapped to auxiliary flat space using the Boyer--Lindquist-like coordinate radius $r$. This numerical kludge prescription has been used in Kerr~\cite{Babak:2006uv}, dCS~\cite{Sopuerta:2009iy}, and other backgrounds in modified gravity~\cite{Lu:2025xlp}. Both backgrounds use the same normalization of coordinate time at infinity, time origin, and observer basis. Orbital periods and harmonic frequencies follow from the geodesic motion, while waveform amplitudes and residual differences depend on the auxiliary mapping~\cite{Berry:2012im}.

The waveform model isolates the conservative orbital contribution; radiation reaction, scalar radiation, and direct modifications to the radiative field are outside the present construction.

The radiation formula at leading quadrupole order relates the emitted field to the second time derivative of the mass quadrupole~\cite{Flanagan:2005yc}. For a source of physical mass $\mu$, the symmetric trace-free (STF) mass quadrupole $I_{ij}=\mu[X_iX_j]^{\rm STF}$ gives the radiation tensor before transverse polarization projection,
\bqn
h_{ij}&=&\frac{2G\mu}{c^4D_L}
\left[A_iX_j+A_jX_i+2V_iV_j\right]^{\rm STF}. \label{sec5:radiationtensor}
\eqn
Here $D_L$ is the physical luminosity distance; derivatives use coordinate time, with a common propagation delay absorbed into its origin. For observer inclination $\iota$ and azimuth $\beta$, the transverse directions are
\bqn
\boldsymbol e_\iota&=&(\cos\iota\cos\beta,\cos\iota\sin\beta,-\sin\iota), \nb\\
\boldsymbol e_\beta&=&(-\sin\beta,\cos\beta,0). \label{sec5:observerbasis}
\eqn
The observer components are $h_{\iota\iota}=e_\iota^ie_\iota^jh_{ij}$, $h_{\beta\beta}=e_\beta^ie_\beta^jh_{ij}$, and $h_{\iota\beta}=e_\iota^ie_\beta^jh_{ij}$~\cite{Zhang:2025wni}. The polarizations in our common Kerr--dCS convention are
\bqn
h_+&=&\dfrac{1}{2}(h_{\iota\iota}-h_{\beta\beta}),\qquad
h_\times=h_{\iota\beta}. \label{sec5:polarizations}
\eqn

The orbit calculation uses $G=c=M=1$. Bars denote dimensionless numerical quantities, including the closure period $\bar T_{\rm cl}$. Physical positions, time, and period are restored by
\bqn
X^i&=&\frac{GM}{c^2}\bar X^i,\qquad t=t_g\bar t, \nb\\
T_{\rm cl}(M;\zeta)&=&t_g\bar T_{\rm cl}(\zeta),\qquad t_g=\frac{GM}{c^3}.
\label{sec5:massunits}
\eqn
The normalized closure time is $u=t/T_{\rm cl}=\bar t/\bar T_{\rm cl}$. The same waveform expressed in this variable is $H_A(u;\zeta)\equiv h_A[uT_{\rm cl}(M;\zeta);M,\zeta]$, where $A=+,\times$. For the equatorial orbit at $\beta=0$, substituting Eqs.~\eqref{sec5:embedding}--\eqref{sec5:polarizations} gives
\bqn
\begin{pmatrix}H_+\\H_\times\end{pmatrix}&=&\frac{G\mu}{c^2D_L\bar T_{\rm cl}^{\,2}}
\frac{d^2}{du^2}
\begin{pmatrix}
\cos^2\!\iota\,\bar X^2-\bar Y^2\\
2\cos\iota\,\bar X\bar Y
\end{pmatrix}. \label{sec5:scaledstrain}
\eqn
Here $(\bar X,\bar Y)$ are the in-plane components of the dimensionless trajectory. With $d/dt=(t_g\bar T_{\rm cl})^{-1}d/du$, the mass dependence of the two length factors in the quadrupole cancels that of the two time derivatives at fixed dimensionless orbit and physical $\mu,D_L$. The physical period remains proportional to $M$.

The dCS signal and Kerr reference share the physical mass $M_{\rm ref}$, source and observer parameters, time origin, and initial azimuth. All coupling derivatives follow the matched family at fixed $(a,L,q_*)$, including its energy adjustment. At fixed $z$ and $M_{\rm ref}$, the period coefficient $C_T=\delta T_r/T_r^{\rm K}$ [Eq.~\eqref{sec4:period_split}] also equals $\delta\bar T_{\rm cl}/\bar T_{\rm cl}^{\rm K}$.

At fixed $u$, Eq.~\eqref{sec5:scaledstrain} gives $H_A(u;\zeta)=H_A^{\rm K}(u)+\zeta\delta H_A(u)+\mathcal O(\zeta^2)$, defining the first-order coefficient $\delta H_A$. It includes the trajectory deformation and $-2C_TH_A^{\rm K}$ from the explicit $\bar T_{\rm cl}^{-2}$ factor, since $[\bar T_{\rm cl}^{\rm K}/\bar T_{\rm cl}(\zeta)]^2=1-2\zeta C_T+\mathcal O(\zeta^2)$.

A comparison at the same physical time $t$ uses each waveform's own normalized time: $u(t;\zeta)=t/T_{\rm cl}(M_{\rm ref};\zeta)$ for dCS and $u_{\rm K}=t/T_{\rm cl}^{\rm K}(M_{\rm ref})$ for Kerr. The period expansion gives
\bqn
\frac{T_{\rm cl}(M_{\rm ref};\zeta)}{T_{\rm cl}^{\rm K}(M_{\rm ref})}
&=&1+\zeta C_T+\mathcal O(\zeta^2), \\
u(t;\zeta)&=&u_{\rm K}\bigl[1-\zeta C_T+\mathcal O(\zeta^2)\bigr]. 
\eqn

Let $\left.\delta h_A\right|_t$ denote the coupling derivative at $\zeta=0$ with the physical time $t$ held fixed. Differentiating $h_A(t;M_{\rm ref},\zeta)=H_A[u(t;\zeta);\zeta]$ and using the expansion above gives
\bqn
\left.\delta h_A\right|_t&=&\delta H_A(u_{\rm K})-C_Tu_{\rm K}(H_A^{\rm K})'(u_{\rm K}). \label{sec5:clock}
\eqn

A prime denotes $d/du$. The first term is the complete correction at fixed $u$; the second arises only from the change of waveform argument at fixed $t$ and is the period-induced phase correction. The finite first-order corrections are $\zeta\delta H_A$ and $\zeta\left.\delta h_A\right|_t$. The variable $u_{\rm K}$ counts all elapsed Kerr closure cycles, with the periodic waveform functions extended over successive cycles.

\begin{figure*}[!t]
\centering
\includegraphics[width=0.88\textwidth]{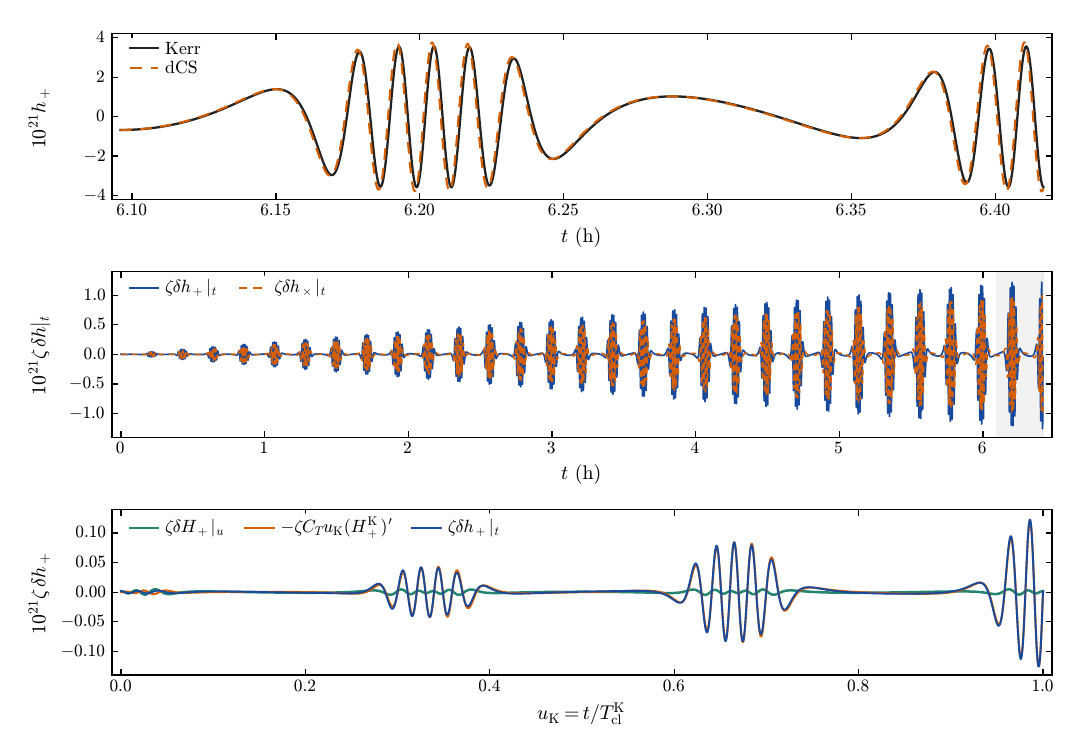}
\caption{Quadrupole waveforms for the baseline prograde $(3,2,1)$ orbit at $a=0.99$, $\zeta=10^{-5}$, and equal physical mass $M_{\rm ref}$. Top: Kerr and first-order dCS $+$ waveforms during the final half of the tenth closure. Middle: both polarization corrections over ten closures at fixed physical time; shading selects the upper interval. Bottom: the $+$ correction over the first closure, split into contributions at fixed $u$ and from the period-induced phase shift [Eq.~\eqref{sec5:clock}]. The term at fixed $u$ also gives the residual after mass matching [Eqs.~\eqref{sec5:waveform_mass_difference} and \eqref{sec5:massmatch}].}
\label{fig6_waveforms}
\end{figure*}

\subsection{Phase accumulation, mass matching, and \texorpdfstring{$C_T=0$}{CT=0} orbits}\label{sec5:clocks}

Figure~\ref{fig6_waveforms} uses the $(3,2,1)$ orbit at $a=0.99$, $L=2.0589192302$, and $\zeta=10^{-5}$. The physical parameters are $M=M_{\rm ref}=10^6M_\odot$, $\mu=10M_\odot$, $D_L=100\,\mathrm{Mpc}$, $\iota=\pi/3$, and $\beta=0$. The reference physical period is $T_{\rm cl}^{\rm K}(M_{\rm ref})=2310.05\,\mathrm{s}$, so ten closures span $6.4168\,\mathrm{h}$.

The upper plot of Fig.~\ref{fig6_waveforms} shows the peaks separating at equal physical times, while the middle plot shows the corrections developing a growing oscillatory envelope. The lower plot identifies its origin: the periodic term $\delta H_A(u_{\rm K})$ remains bounded, whereas the period-induced phase correction grows with $u_{\rm K}$. We quantify these differences using the relative time-domain $L^2$ norm employed in analyses of waveform accuracy~\cite{Lindblom:2010mh}. Both comparisons use $0\le t\le T$ and equal weights for the two polarizations, without noise weighting:

\bqn
\epsilon_t(T)&=&\zeta\sqrt{
\frac{\displaystyle\sum_{A=+,\times}\int_0^T
\bigl(\left.\delta h_A\right|_t\bigr)^2\,dt}
{\displaystyle\sum_{A=+,\times}\int_0^T
\bigl[h_A^{\rm K}(t)\bigr]^2\,dt}}, \nb\\[6pt]
\epsilon_u(T)&=&\zeta\sqrt{
\frac{\displaystyle\sum_{A=+,\times}\int_0^T
\bigl[\delta H_A(u_{\rm K})\bigr]^2\,dt}
{\displaystyle\sum_{A=+,\times}\int_0^T
\bigl[h_A^{\rm K}(t)\bigr]^2\,dt}}. \label{sec5:l2}
\eqn

Over ten closures, the relative norms of the first-order waveform corrections defined in Eq.~\eqref{sec5:l2} are $15.30\%$ at fixed physical time and $0.1081\%$ at equal normalized closure time. The larger norm at fixed physical time is dominated by the phase accumulated through the period shift.

The common frequency shift from Eq.~\eqref{sec3w:frequency_relation} accumulates a phase difference in harmonic $k$,

\bqn
\Delta\Phi_k(T)&=&2\pi\zeta\delta f_kT,
\qquad \delta f_k=-C_Tf_k^{\rm K}. \label{sec5:phaseacc}
\eqn

For the dominant Kerr harmonic $k=47$, the phase difference reaches approximately $0.31\,\mathrm{rad}$ over ten closures. This linearized comparison describes conservative periodic motion while the accumulated first-order phase remains small.

A nearby Kerr mass can reproduce the period-induced phase correction. We keep the dimensionless orbit $(a,L,q_*)$, the physical mass $\mu$ of the small body, distance, viewing angles, and initial time and phase fixed. The waveform amplitude in Eq.~\eqref{sec5:scaledstrain} is then independent of $M$. At fixed physical time, the template's normalized closure time is $t/T_{\rm cl}^{\rm K}(M)=c^3t/(GM\bar T_{\rm cl}^{\rm K})$. Its mass derivative at $M=M_{\rm ref}$ is $-u_{\rm K}/M_{\rm ref}$, so the chain rule gives

\bqn
\left.\frac{\partial h_A^{\rm K}(t;M)}{\partial M}\right|_{\substack{t\\M=M_{\rm ref}}}
&=&-\frac{u_{\rm K}}{M_{\rm ref}}(H_A^{\rm K})'(u_{\rm K}).
\label{sec5:massdirection}
\eqn

Let $M_{\rm K}$ denote the mass of the nearby Kerr template, with $(M_{\rm K}-M_{\rm ref})/M_{\rm ref}=\mathcal O(\zeta)$. Since $T_{\rm cl}^{\rm K}(M_{\rm K})/T_{\rm cl}^{\rm K}(M_{\rm ref})=M_{\rm K}/M_{\rm ref}$, expanding at the same physical time gives
\bqn
h_A^{\rm K}(t;M_{\rm K})&=&H_A^{\rm K}\!\left(u_{\rm K}\frac{M_{\rm ref}}{M_{\rm K}}\right) \nb\\
&=&H_A^{\rm K}(u_{\rm K})-\frac{M_{\rm K}-M_{\rm ref}}{M_{\rm ref}}\,
u_{\rm K}(H_A^{\rm K})'(u_{\rm K}) \nb\\
&&+\mathcal O(\zeta^2). \nb
\eqn
Combining this expansion with Eq.~\eqref{sec5:clock} gives the waveform difference at a nearby Kerr template mass:
\bqn
&&h_A^{\rm dCS}(t;M_{\rm ref})-h_A^{\rm K}(t;M_{\rm K}) \nb\\
&&\quad=\zeta\delta H_A(u_{\rm K}) \nb\\
&&\qquad+\left[\frac{M_{\rm K}-M_{\rm ref}}{M_{\rm ref}}-\zeta C_T\right]
u_{\rm K}(H_A^{\rm K})'(u_{\rm K}) \nb\\
&&\qquad+\mathcal O(\zeta^2). \label{sec5:waveform_mass_difference}
\eqn
Equating the coefficients of $u_{\rm K}(H_A^{\rm K})^{\prime}$ gives the Kerr mass required to match the orbital periods:
\bqn
\frac{M_{\rm K}-M_{\rm ref}}{M_{\rm ref}}&=&\zeta C_T+\mathcal O(\zeta^2).
\label{sec5:massmatch}
\eqn

For $C_T=-10.5186$, $M_{\rm ref}=10^6M_\odot$, and $\zeta=10^{-5}$, Eq.~\eqref{sec5:massmatch} gives $M_{\rm K}-M_{\rm ref}\simeq-105.19M_\odot$. This mass adjustment cancels the coefficient of $u_{\rm K}(H_A^{\rm K})'$ in Eq.~\eqref{sec5:waveform_mass_difference}, leaving $\zeta\delta H_A(u_{\rm K})$.

For $C_T=0$, Eq.~\eqref{sec5:waveform_mass_difference} gives the residual $\zeta\delta H_A$ at $M_{\rm K}=M_{\rm ref}$. Mass matching varies the Kerr template mass at a fixed dimensionless orbit. At a different reference orbit with $C_T=0$, the first-order period correction vanishes through cancellation between the direct contribution and energy compensation. Both comparisons isolate the full periodic correction defined in Eq.~\eqref{sec5:scaledstrain}.

Related degeneracies in frequency matching have been studied by varying orbital parameters in quasi-Kerr models~\cite{Glampedakis:2005cf} and the Kerr mass and angular momentum in black-hole--torus models~\cite{Barausse:2006vt}.

The upper plot of Fig.~\ref{fig8_cancellation_waveform_spectra} shows the waveform components of the $C_T=0$ orbit. At $a=0.99$ and $f_L\simeq0.404534$ on the prograde $C_T=0$ boundary in Sec.~\ref{section4}, the $(3,2,1)$ orbit has \mbox{$C_T^{\mathrm{dir}}\simeq-C_T^{\mathrm{comp}}\simeq3.732733$}. We use the unrounded zero in the waveform calculation; its numerical stability is checked in Sec.~\ref{section3}. Its nonzero waveform correction gives $\epsilon_t=\epsilon_u=0.08016\%$ at $\zeta=10^{-5}$. The curve at fixed $u$ in the lower plot of Fig.~\ref{fig6_waveforms} gives the baseline residual after the mass adjustment; the corresponding curve in the upper plot of Fig.~\ref{fig8_cancellation_waveform_spectra} gives the residual of the $C_T=0$ orbit at equal mass.

\begin{figure*}[!t]
\centering
\includegraphics[width=\textwidth]{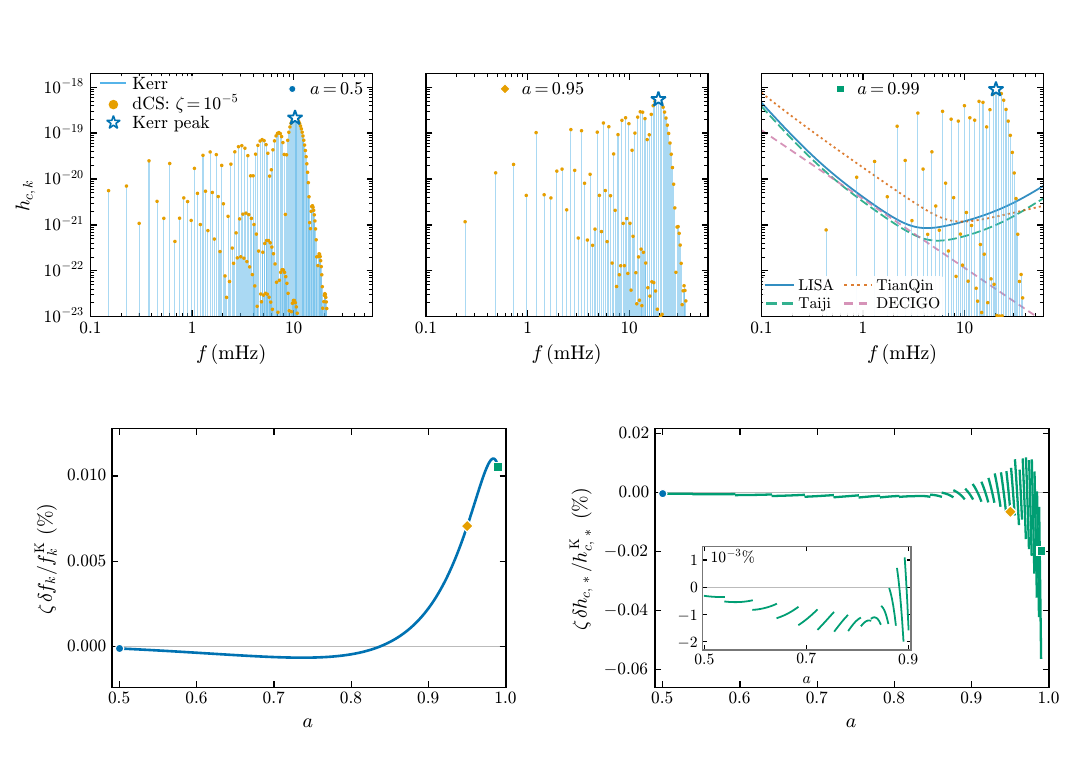}
\caption{Periodic spectra and first-order corrections for the baseline prograde $(3,2,1)$ family at $\zeta=10^{-5}$ and equal Kerr and dCS physical masses. Top: characteristic strains $h_{c,k}$ at $a=0.50,0.95,0.99$ (left to right), with light-blue Kerr stems, orange dCS markers, and coherent normalization $T_{\rm ref}=1\,\mathrm{yr}$. Stars mark Kerr peaks $k_*=137,80,47$, respectively. Noise references for LISA, Taiji, TianQin, and DECIGO appear at upper right. Bottom: the common fractional frequency shift $\zeta\delta f_k/f_k^{\rm K}=-\zeta C_T$ (left) and relative strain correction $\zeta\delta h_{c,*}/h_{c,*}^{\rm K}$ for the strongest Kerr harmonic at each spin (right), in percent. Breaks mark changes in $k_*$. Circles, diamonds, and squares mark the three representative spins. Other source and observer parameters are given in Sec.~\ref{sec5:clocks}.}
\label{fig7_spectra}
\end{figure*}

\begin{figure*}[!t]
\centering
\includegraphics[width=0.96\textwidth]{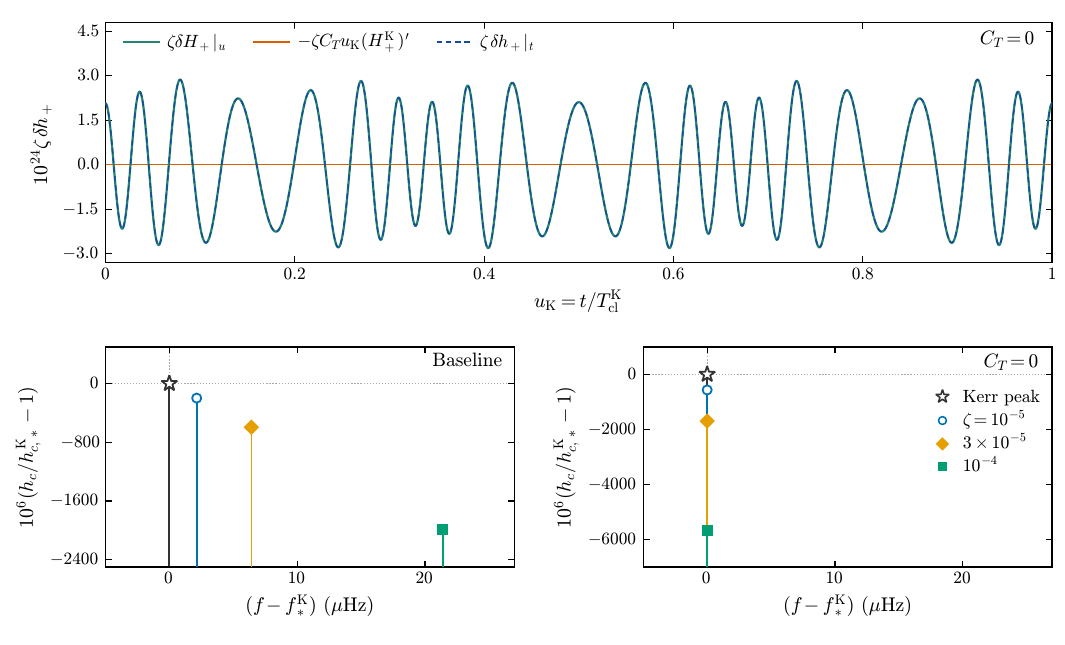}
\caption{Waveform decomposition and peak responses for two prograde $(3,2,1)$ orbits at $a=0.99$: the baseline $f_L=0.776642$, $C_T=-10.5186$, and $C_T=0$ at $f_L\simeq0.404534$. Each has its own Kerr reference at fixed $(a,L,q_*)$ and equal physical mass $M_{\rm ref}$. Top ($C_T=0$): contributions at fixed $u$, from the period-induced phase shift, and their sum for the $+$ correction at $\zeta=10^{-5}$ over the first Kerr closure, with $u_{\rm K}=t/T_{\rm cl}^{\rm K}$. The phase term vanishes, leaving the residual at fixed $u$. Bottom: Kerr peak harmonics $k_*=47$ (baseline, left) and $23$ ($C_T=0$, right) at $\zeta=10^{-5},3\times10^{-5},10^{-4}$, with frequency offsets in $\mu\mathrm{Hz}$ and relative characteristic strain changes in ppm. The peak plots share frequency scales and a legend; strain scales differ. Markers rescale the same first-order coefficients. Spectra use $T_{\rm ref}=1\,\mathrm{yr}$. Other parameters are given in Sec.~\ref{sec5:clocks}; the baseline decomposition appears in Fig.~\ref{fig6_waveforms}.}
\label{fig8_cancellation_waveform_spectra}
\end{figure*}

\subsection{Periodic spectra and their spin dependence}\label{sec5:spectra}

A closed orbit has a discrete spectrum. For positive integers $k$, we define the Fourier coefficients, harmonic frequencies, and combined line amplitudes by
\bqn
c_{A,k}&=&\int_0^1 H_A(u)e^{-2\pi iku}\,du, \nb\\*
f_k&=&\frac{k}{T_{\rm cl}},\qquad
\mathcal A_k=2\left(\sum_{A=+,\times}|c_{A,k}|^2\right)^{1/2}.
\label{sec5:amplitude}
\eqn
The first-order Fourier coefficients and, for harmonics with $\mathcal A_k^{\rm K}\ne0$, the corresponding amplitude corrections follow from Eq.~\eqref{sec5:amplitude}:
\bqn
\delta c_{A,k}&=&\int_0^1\delta H_A(u)e^{-2\pi iku}\,du, \nb\\
\delta\mathcal A_k&=&\frac{4}{\mathcal A_k^{\rm K}}\operatorname{Re}
\sum_{A=+,\times}(c_{A,k}^{\rm K})^*\delta c_{A,k}. \nb
\eqn
Here $\operatorname{Re}$ takes the real part and the asterisk denotes complex conjugation.

For an ideal stationary periodic signal, we use the coherent reference duration $T_{\rm ref}=1\,\mathrm{yr}$ (365.25 days) for spectral normalization; the time-domain comparison covers ten Kerr closures. The continuous characteristic strain $h_c(f)=2f|\widetilde h(f)|$ and its weighting in logarithmic frequency are given in Ref.~\cite{Moore:2014lga}. Reference~\cite{Robson:2018ifk} discusses a definition of characteristic strain that combines both polarizations. It also describes a point representation based on the signal-to-noise ratio for slowly evolving sources. Here $\widetilde h_A(f)=\int_0^{T_{\rm ref}}h_A(t)e^{-2\pi ift}\,dt$. For resolved harmonics with $T_{\rm ref}\gg T_{\rm cl}$, Parseval's identity gives the characteristic strain of a discrete spectral line through
\bqn
h_{c,k}^{\,2}&\equiv&4f_k\sum_{A=+,\times}\int_{\mathrm{line}\ k}
|\widetilde h_A(f)|^2\,df \nb\\
&\simeq&4f_kT_{\rm ref}\sum_{A=+,\times}|c_{A,k}|^2
=\mathcal A_k^2 f_kT_{\rm ref}. \nb
\eqn
In the narrow-line limit, $h_{c,k}=\mathcal A_k\sqrt{f_kT_{\rm ref}}$ is the square root of the integral over logarithmic frequency of the squared continuous characteristic strain across the line. At fixed $k,T_{\rm ref}$, the product rule and $\delta f_k/f_k^{\rm K}=-C_T$ [Eq.~\eqref{sec3w:frequency_relation}] give
\bqn
\delta h_{c,k}&=&\sqrt{f_k^{\rm K}T_{\rm ref}}
\left(\delta\mathcal A_k-\dfrac{1}{2}C_T\mathcal A_k^{\rm K}\right).
\label{sec5:hc}
\eqn

At fixed dimensionless Kerr orbit and physical $\mu,D_L$, $\mathcal A_k^{\rm K}$ is independent of the template mass, whereas $f_k^{\rm K}\propto M_{\rm K}^{-1}$. Thus $h_{c,k}^{\rm K}\propto M_{\rm K}^{-1/2}$ for fixed $T_{\rm ref}$. After the mass adjustment in Eq.~\eqref{sec5:massmatch}, the explicit contribution from the frequency factor in Eq.~\eqref{sec5:hc} cancels against the corresponding change of the Kerr template. The remaining difference in characteristic strain is $\zeta\sqrt{f_k^{\rm K}T_{\rm ref}}\,\delta\mathcal A_k+\mathcal O(\zeta^2)$, with $f_k^{\rm K}$ evaluated at $M_{\rm ref}$. The equal-mass $C_T=0$ comparison has the same form, although $\delta\mathcal A_k$ depends on the chosen reference orbit.

For context on frequency and strain scales, the upper-right plot of Fig.~\ref{fig7_spectra} includes the instrumental reference curves of LISA, Taiji, TianQin, and DECIGO~\cite{Robson:2018ifk,Liu:2023qap,Huang:2020rjf,Yagi:2011wg}, using the definition of characteristic noise in Eq.~(18) of Ref.~\cite{Moore:2014lga}:
\bqn
h_n(f)&=&\sqrt{fS_n(f)}.
\label{sec5:noise_reference}
\eqn
Here $S_n$ is the one-sided effective PSD of strain noise. The curves use instrumental responses averaged over sky position and polarization in a convention with two effective channels, excluding Galactic confusion foregrounds; the source inclination remains fixed. The dashed DECIGO curve shows the reference instrumental noise model below its principal $0.1$--$10\,\mathrm{Hz}$ design band~\cite{Kawamura:2020pcg}. Comparison with $h_n(f_k)$ assumes coherence over $T_{\rm ref}$, resolved lines, and noise that varies negligibly across each line, with consistent averaged response and channel conventions~\cite{Robson:2018ifk}.

The absolute spectra nearly overlap because the shifts at $\zeta=10^{-5}$ are small on the logarithmic scales in the upper row of Fig.~\ref{fig7_spectra}. The lower row of Fig.~\ref{fig8_cancellation_waveform_spectra} resolves the two Kerr peak corrections. At first order,
\bqn
f_*(\zeta)-f_*^{\rm K}&=&\zeta\,\delta f_*, \nb\\
\frac{h_{c,*}(\zeta)}{h_{c,*}^{\rm K}}-1&=&\zeta\,\frac{\delta h_{c,*}}{h_{c,*}^{\rm K}}. \label{sec5:peak_offsets}
\eqn
The markers at $\zeta=10^{-5},3\times10^{-5},10^{-4}$ rescale the same first-order coefficients.

The lower-left plot of Fig.~\ref{fig7_spectra} shows the common frequency shift $\zeta\,\delta f_k/f_k^{\rm K}=-\zeta C_T$ along the baseline spin sequence [Eq.~\eqref{sec3w:frequency_relation}]. It changes from slightly negative values at moderate spin to positive values toward high spin, with sign opposite to that of the $(3,2,1)$ period curve in Fig.~\ref{fig2_spin_corrections}. The lower-right plot tracks the first-order characteristic strain correction for $k_*(a)=\arg\max_k h_{c,k}^{\rm K}(a)$, selected independently at each spin. The plotted branches end where $k_*$ changes, distinguishing a change of the tracked line from a continuous variation of its correction.

The dominant Kerr line at $a=0.99$ has $k_*=47$ and $f_*^{\rm K}=20.3458\,\mathrm{mHz}$. For $\zeta=10^{-5}$, the fractional frequency shift is $\zeta\,\delta f_*/f_*^{\rm K}=0.01052\%$, corresponding to $2.1401\,\mu\mathrm{Hz}$, and the fractional change in $h_{c,*}$ is $-0.01985\%$; at $\zeta=10^{-4}$, the linear predictions are $0.10519\%$ ($21.401\,\mu\mathrm{Hz}$) and $-0.19849\%$, respectively. Both the frequency shift and the phase drift dominating the time-domain waveform difference are controlled by $C_T$. A corresponding Kerr mass change reproduces this phase drift, while differences in waveform shape and amplitude remain at equal normalized closure time.

For the $C_T=0$ orbit in Sec.~\ref{sec5:clocks}, all fixed harmonics have zero first-order frequency offset, while the Kerr peak at $k_*=23$ changes in height by $-0.05660\%$ at $\zeta=10^{-5}$ [the lower-right plot of Fig.~\ref{fig8_cancellation_waveform_spectra}]. The two peak plots use their respective Kerr reference harmonics.

\section{Conclusions and outlook}\label{sec:conclusions}

Using the analytic fit to the leading dCS metric correction, we have calculated first-order corrections to the energy, radial period, and circumferential turning radii of three prograde equatorial periodic families over $0.5\leq a\leq0.99$. Each Kerr--dCS pair fixes the dimensionless spin $a$, dimensionless specific angular momentum $L$, and rational periodicity $q_*$.

The period correction is controlled by competition between the direct geometric contribution at fixed energy and the energy adjustment preserving periodicity. For the baseline $(3,2,1)$ orbit at $a=0.99$, the direct contribution is \mbox{$C_T^{\mathrm{dir}}\simeq4.7151$}, while energy compensation gives \mbox{$C_T^{\mathrm{comp}}\simeq-15.2337$}. Their sum is $C_T\simeq-10.5186$. The sign depends on both the periodic family and the reference path.

The two-parameter $(3,2,1)$ scan over $0.01\leq a\leq0.99$ locates $C_T=0$ boundaries and identifies different mechanisms among the sampled orbits with negative $C_T$: energy compensation dominates for prograde motion, whereas the direct contribution dominates for retrograde motion. Circumferential radii further show that geometric turning point changes need not share the signs of coordinate shifts.

Within the leading quadrupole numerical kludge model, the period correction produces a common fractional shift of the harmonic frequencies and an accumulated phase drift. At fixed dimensionless orbit and physical mass $\mu$ of the small body, a Kerr mass adjustment satisfying $(M_{\rm K}-M_{\rm ref})/M_{\rm ref}=\zeta C_T$ reproduces this frequency and phase contribution to first order. The representative $C_T=0$ orbit suppresses the same contribution at equal mass. Both computed comparisons retain nonzero periodic waveform and spectral amplitude corrections.

Including radiation reaction through the tensor and scalar fluxes of dCS gravity would connect the conservative orbital corrections to evolving extreme-mass-ratio inspirals~\cite{Sopuerta:2009iy}. This extension would allow changes in the orbital frequencies to be followed during inspiral and related to the accumulated waveform phase. Generalizing the matching prescription to regular inclined bound motion would also allow the radial, polar, and azimuthal frequencies to be studied together, including resonances encountered during inspiral~\cite{Flanagan:2010cd}.

The approach also applies to smooth branches of equatorial bound orbits in other perturbatively deformed Kerr backgrounds. Perturbative Kerr deformations and rotating solutions in cubic and quartic gravity provide relevant examples~\cite{Cano:2019ore,Burger:2019wkq,Adair:2020vso,Marks:2023ipa}. Recent numerical solutions and analytic metrics also make rapidly rotating cases accessible~\cite{Fernandes:2025vxg,Hu:2026ute}. Comparing the first-order energy and period corrections across these backgrounds could clarify which features are shared and which depend on the curvature interaction. Such comparisons could also test the spin dependence near extremality, subject to the regime of validity of effective field theory~\cite{Lam:2026grz}.

\section*{Acknowledgements}

This work is supported in part by the National Natural Science Foundation of China under Grant Nos. 12675080, 12542053, 12275238 and 11675143, the National Key Research and Development Program under Grant No. 2020YFC2201503, and the Zhejiang Provincial Natural Science Foundation of China under Grant Nos. LR21A050001 and LY20A050002, and the Fundamental Research Funds for the Provincial Universities of Zhejiang in China under Grant No. RF-A2019015. X. Zhang and J.-F. Zhang are supported by the National Natural Science Foundation of China (Grant Nos. 12533001, 12575049, 12473001, and 12447105), the National SKA Program of China (Grant Nos. 2022SKA0110200 and 2022SKA0110203), the China Manned Space Program (Grant No. CMS-CSST-2025-A02), and the 111 Project (Grant No. B16009). OpenAI's ChatGPT and Codex [GPT-6 Astra] were used to improve the language of the manuscript, help cross-check algebraic calculations and derivations, and assist with code modification and numerical checks. The authors take full responsibility for the scientific content and conclusions of this work.

\FloatBarrier

\end{document}